# Large negative magnetoresistance in thiospinel CuCrZrS$_4$

Takao Furubayashi [a,*], Hiroyuki Suzuki [a], Nami Kobayashi [b], and Shoichi Nagata [b]

[a] Nanomaterials Laboratories, National Institute for Materials Science,

1-2-1 Sengen, Tsukuba 305-0047, Japan

[b] Department of Materials Science and Engineering, Muroran Institute of Technology,

27-1 Mizumoto-cho, Muroran 050-8585, Japan

Abstract

We report on large negative magnetoresistance observed in ferromagnetic thiospinel compound CuCrZrS$_4$. Electrical resistivity increased with decreasing temperature according to the form proportional to $\exp(T_0/T)^{1/2}$, derived from variable range hopping with strong electron-electron interaction. Resistivity under magnetic fields was expressed by the same form with the characteristic temperature $T_0$ decreasing with increasing magnetic field. Magnetoresistance ratio $\rho(T,0)/\rho(T,H)$ is 1.5 at 100 K for $H$ = 90 kOe and increases divergently with decreasing temperature reaching 80 at 16 K. Results of magnetization measurements are also presented. Possible mechanism of the large magnetoresistance is discussed.

---

[*] Corresponding author.

Tel: +81 29 859 2754, Fax: +81 29 859 2701

E-mail: furubayashi.takao@nims.go.jp

Very large negative magnetoresistance (MR) called colossal magnetoresistance (CMR) in manganite perovskites [1,2] has been extensively studied in view of both physics and applications. It is believed that the phase transition from insulating to metallic induced by magnetic fields is the origin of CMR. Similar CMR effects have been observed in other groups of materials including thiospinel such as $FeCr_2S_4$.[3] It is interesting to notice that large negative MR with no magnetic phase transition has been reported in some magnetic semiconductors such as $Gd_{3-x}S_4$ [4] and amorphous Gd-Si. [5, 6] In these materials, the resistance increases exponentially with decreasing temperature according to variable range hopping (VRH) [7] at low temperatures. Large negative MR is observed at the same time. MR ratio, defined here as $\rho(T,0)/\rho(T,H)$ for the resistivity $\rho$ and the magnetic field $H$, increases divergently with decreasing temperature $T$ and reaches $10^5$ at 1 K for 80kOe in amorphous Gd-Si. [6] No such MR is observed in similar amorphous alloys consisting of non-magnetic components such as Y-Si. Thus, the large MR is believed to result from the formation of magnetic polarons, which originate from the scattering of conduction electrons with the localized large magnetic moment of Gd.

This paper describes large negative MR combined with VRH conduction observed in sulfospinel CuCrZrS$_4$. Spinel $CuCr_2S_4$ is a well known metallic and ferromagnetic compound with the Curie temperature $T_C$ of about 400 K. [8-11] Recently, spinel CuCrZrS$_4$ was obtained by substituting half of Cr by Zr. [12] It has the same normal spinel structure with Cu in the *A* site, which is tetrahedrally coordinated with S, and Cr and Zr in the *B* site, which is octahedrally coordinated. By this substitution, $T_C$ is decreased to about 60 K. In addition, a successive magnetic transition suggesting reentrant spin glass was observed at the temperature $T_G \approx 10$ K. As for the electrical properties, semiconducting behavior of the resistivity was observed. In this work, we extended the studies of magnetization and electrical resistance to the region of high magnetic fields. The most striking result is that large negative MR is observed. MR ratio for $H$ = 90 kOe reaches 80 at 16 K. We show that the MR is closely related to VRH conduction in this material.

Powder samples were prepared by the solid reaction method as described previously. [12] At the final stage of the preparation, the powder specimens were reground, pressed into rectangular bars and then they were sintered at 1023K for 2 days. Powder x-ray diffraction showed the normal spinel structure with the lattice constant $a$ = 1.0171 nm at room temperature. No sign of ordered structure of Cr and Zr was observed: the B sites are randomly occupied by Cr and Zr atoms. The resistivity of the sintered specimens with dimensions of about 2mm×2mm×10mm was measured by a standard dc four-probe method. The dc magnetization was measured with an extraction magnetometer. Both measurements were done at temperatures ranging from 2 K to 300 K under magnetic fields up to $H$ = 90 kOe by using PPMS manufactured by Quantum Design.

Figure 1 shows the magnetization measured at the field of 50 Oe. The Curie temperature, defined here as the temperature corresponding to the peak of the temperature coefficient of the magnetization, was 54 K. The magnetization above $T_c$ is well reproduced by the Curie Weiss behavior, $\chi = C/(T-\Theta) + \chi_0$ with the Weiss temperature $\Theta$ = 53 K. If we assume that all the moment is ascribed to Cr, the paramagnetic moment per Cr atom is obtained to be 3.8$\mu_B$. The value is comparable with 3.87$\mu_B$ calculated by $\mu_{eff} = g\{S(S+1)\}^{1/2}$ with S=3/2 for $Cr^{3+}$. As reported previously [12], the magnetization measured with increasing the temperature after cooled to 2 K in zero magnetic field (ZFC) deviates from that measured with decreasing temperature under the magnetic field (FC). This result suggests a re-entrant spin glass phase below $T_G \approx$10 K. Magnetization curves at various temperatures are shown in Fig. 2. The magnetization does not saturate in magnetic fields up to 90 kOe even at 2 K. The spontaneous magnetization at 2 K, obtained by extrapolating the magnetization curve to 0 Oe, is $\mu_0$=0.3$\mu_B$ per unit formula of $CuCrZrS_4$, namely per Cr atom. The value is significantly smaller than the paramagnetic moment $\mu_{eff}$ = 3.8 $\mu_B$.

The resistivity $\rho$ of the sintered sample was 1.5 Ω·cm at room temperature and increased with decreasing temperature $T$. The fitting in the range from 300 K to 100 K to the activation-type form $\rho = A\exp(\Delta/k_BT)$ gives the activation energy, $\Delta/k_B \approx 300$ K. The temperature dependence deviates from the activation type at low temperatures below about 100 K. As shown in Fig. 3, it is well reproduced by the form $\rho = \rho_0 \exp(T_0/T)^\alpha$ with α=1/2, derived from VRH with strong electron-electron interaction. [13, 14] The same form with α=1/4, for VRH in a three dimensional system without electron-electron interaction, was also tried but resulted in a narrower temperature range of good fitting.

The resistivity measured with magnetic fields applied in the direction parallel to the current is shown in Fig. 4. The direction of the field to the current had no significant influence: MR is always isotropic and negative. The temperature dependence of $\rho$ under magnetic fields is also well expressed by the same form $\rho = \rho_0 \exp(T_0/T)^{1/2}$ as in zero magnetic field. The characteristic temperature $T_0$ is found to decrease with increasing the field as shown in the inset of Fig.4, similarly as observed in amorphous Gd-Si. [6] Theoretically, $T_0$ is expressed by $k_BT_0 = 2.8e^2/\kappa\xi$, where $\kappa$ is the dielectric constant and $\xi$ the localization length. [13, 14] Thus, it appears that $\kappa$ and/or $\xi$ are enhanced by magnetic fields. Figure 5 describes the field dependence at each temperature. MR ratio $\rho(T, 0)/\rho(T, H)$ for $H = 90$ kOe is 1.5 at 100 K and increased with decreasing temperature to 80 at 16 K.

For the electronic state of $CuCr_2S_4$, Lotgering [8] proposed the ionic model $Cu^{1+}Cr^{3+}Cr^{4+}S^{2-}_4$. Recent x-ray spectroscopic studies have shown that the magnetic moment on the Cu site is very small. [11] The result is consistent with the ionic state $Cu^{1+}$ with the $3d$ states fully occupied. The metallic conduction of the material is explained as the result of the mixed valence state of $Cr^{3+}$ and $Cr^{4+}$. When half of the Cr atoms are replaced by Zr, two ionic states $Zr^{3+}$ and $Zr^{4+}$ are probable. Considering higher ionization tendency of Zr than Cr, however, $Zr^{4+}$ would be preferable to $Zr^{3+}$. The ionic configuration would be $Cu^{1+}Cr^{3+}Zr^{4+}S^{2-}_4$. Then, it is reasonable to consider that the electrical conduction is mostly determined by the hopping of $3d$ electrons localized at $Cr^{3+}$ because $3d$ orbitals of both $Cu^{1+}$ and $Zr^{4+}$ are fully occupied. Electron transfer from $Cr^{3+}$ to neighboring $Cr^{3+}$ would require some excess energy due to the on-site Coulomb energy $U$. In addition, Cr and Zr atoms randomly distributed on the $B$ site would form a random potential for electrons. The combination of $U$ and the random potential leads to VRH with strong electron-electron interaction. Thus, the ionic model is consistent with the observed VRH conduction at low temperatures.

From the Curie-Weiss paramagnetic susceptibility, we obtained $\mu_{eff} = 3.8\mu_B$ per Cr atom, which is comparable with the paramagnetic moment derived from $(3d)^3$ of $Cr^{3+}$. Thus, the paramagnetic behavior above $T_C$ is consistent with the ionic model. On the other hand, the magnetic structure below $T_C$ is still controversial. The positive $\Theta$ comparable with $T_C$ may suggest a ferromagnetic ordering. However, the small spontaneous magnetization and the non-saturating magnetization curves seem contradicting with the ferromagnetic picture. The presence of some non-collinear spin structure possibly explains the results. It is to be noted that, although the positive $\Theta$ indicates that the exchange interaction is ferromagnetic on average, the presence of antiferromagnetic couplings is not ruled out. One possible magnetic structure is the assembly of ferromagnetic domains that are antiferromagnetically coupled. Helical spin structure would be also possible. However, the detailed magnetic structure is unclear at present and further studies by using microscopic measurements would be required to clarify the problem.

Now we discuss the origin of MR. CMR in manganites or spinel $FeCr_2S_4$ is observed at temperatures around the magnetic phase transition. This is in contrast to the present material. Both the resistivity and the MR ratio increase with cooling with no anomaly at temperatures around $T_C$ as shown in Figs.3, 4 and 5. Thus, the mechanism of CMR is not applicable. Obviously, MR of $CuCrZrS_4$ is closely related to VRH conduction similarly with $Gd_{3-x}S_4$ and amorphous Gd-Si. Scattering of conducting $s$ electrons by localized $4f$ moments of Gd is considered to be the origin of MR. In contrast to these materials, both electronic conduction and magnetic moment seem to exist on the $Cr^{3+}$ sites in $CuCrZrS_4$. Thus, the VRH conduction is considered to occur at $Cr^{3+}$ sites with spin polarization.

One possible explanation for the MR based on VRH is described as the followings. We showed non-saturating behaviors of magnetization curves below $T_C$. Thus, although the magnetic structure is still unclear, both up and down spin sites exist for small magnetic fields. It is plausible that the probability of electron hopping is spin dependent: the probability of electron hopping from an up-spin Cr site to another up-spin Cr site would be larger than that from an up-spin site to a down-spin site. The ratio of the up spin sites increases and the magnetization increases with increasing the magnetic field. Then, the probability of finding up-spin site in the neighbor of an up-spin site increases. As the result, the total hopping probability increases resulting in enhancing the localization length $\xi$. Then, $T_0$ is decreased through the relation $k_B T_0 = 2.8 e^2/\kappa\xi$. The mechanism is somewhat analogous to CMR in manganites, where electron transfer in ferromagnetic state is much enhanced compared to the antiferromagnetic phase. Further theoretical and experimental studies would be required for quantitative evaluation of the model. It is to be noted that MR shown in this paper is observed at temperatures above the re-entrant spin-glass transition temperature, $T_G \approx 10$ K. The resistivity becomes too large at low temperatures around $T_G$ to be measured by our system at present. The influence of spin-glass transition on MR is of much interest and it is to be investigated in future.

To summarize, electronic and magnetic properties of thiospinel $CuCrZrS_4$ were examined in this work. Small spontaneous magnetization and non-saturating magnetization curves below $T_C$ suggest non-collinear spin structure. Temperature dependence of the resistivity is well described by the form $\rho = \rho_0 \exp(T_0/T)^{1/2}$ derived from VRH with strong electron-electron interaction. The resistivity under magnetic fields is expressed by the same form with the characteristic temperature $T_0$ decreasing with increasing the magnetic field. As the result, very large negative MR was observed at low temperatures. No significant influence of the direction of the magnetic field to the current on MR was observed. MR ratio $\rho(T,0)/\rho(T,H)$ increases divergently with decreasing temperature, reaching 80 at 16 K with $H$ = 90 kOe. We suggest a possible explanation that the large MR results from spin-dependent electron hopping.

The authors would like to thank T. Matsumoto, H. Kitazawa and H. Mamiya for fruitful discussions.

Figure Captions

Fig. 1. Magnetization and inverse susceptibility of CuCrZrS$_4$ measured in the field of 50 Oe against the temperature. Field cooled (FC) and zero field cooled (ZFC) branches are shown.

Fig. 2. Magnetization curve measured with decreasing the magnetic field at each temperature. The values are shown per formula unit of CuCrZrS$_4$.

Fig. 3. Logarithm of electrical resistivity plotted against $T^{-1/4}$ and $T^{-1/2}$ for the temperature $T$ ranging from 16K to 300K.

Fig. 4. Electrical resistivity plotted logarithmically against $T^{-1/2}$ under each magnetic field $H$ applied in the direction parallel to the current. The characteristic temperature $T_0$ obtained from the fitting to the form $\rho = \rho_0 \exp(T_0/T)^{1/2}$ is plotted against $H$ in the inset.

Fig. 5. The resistivity normalized by the value at 0 Oe at indicated temperatures. Open circles denote measurements with the current perpendicular to the magnetic field at 16 K. Other points are for the measurements with the current parallel to the magnetic field.

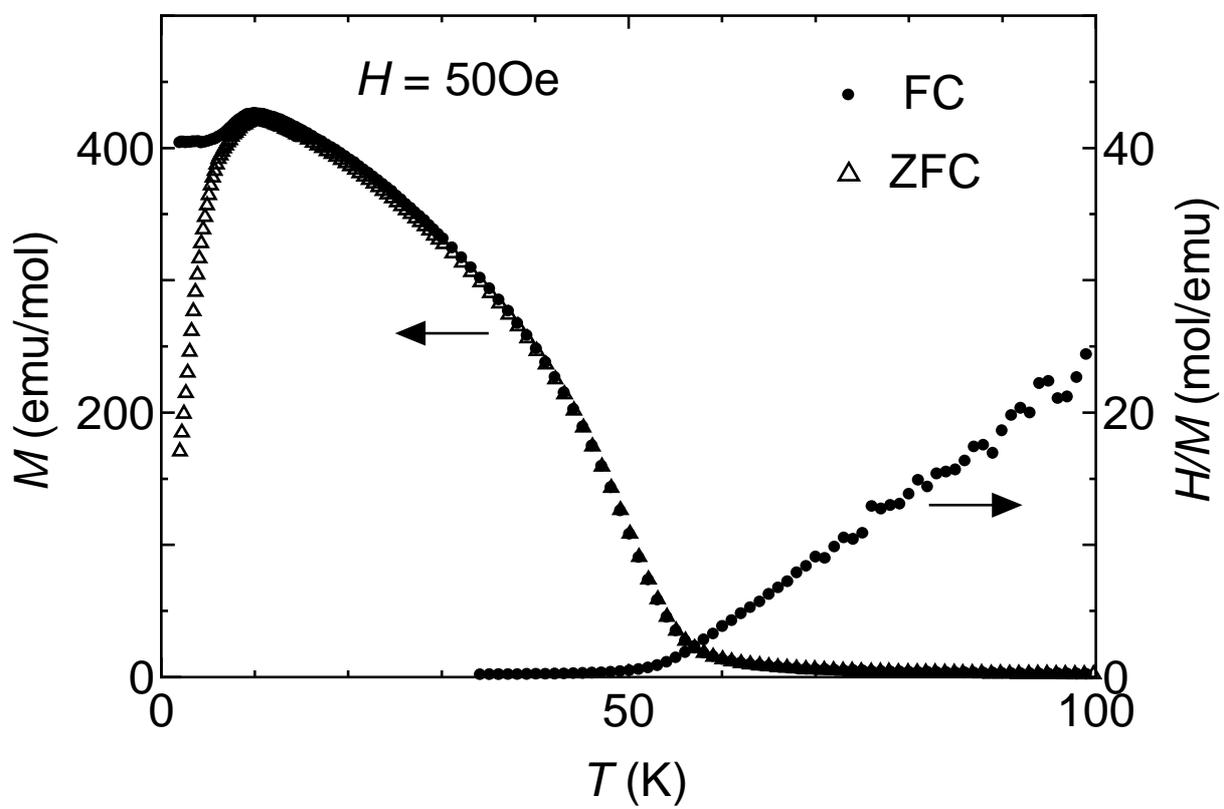

Fig.1, T. Furubayashi

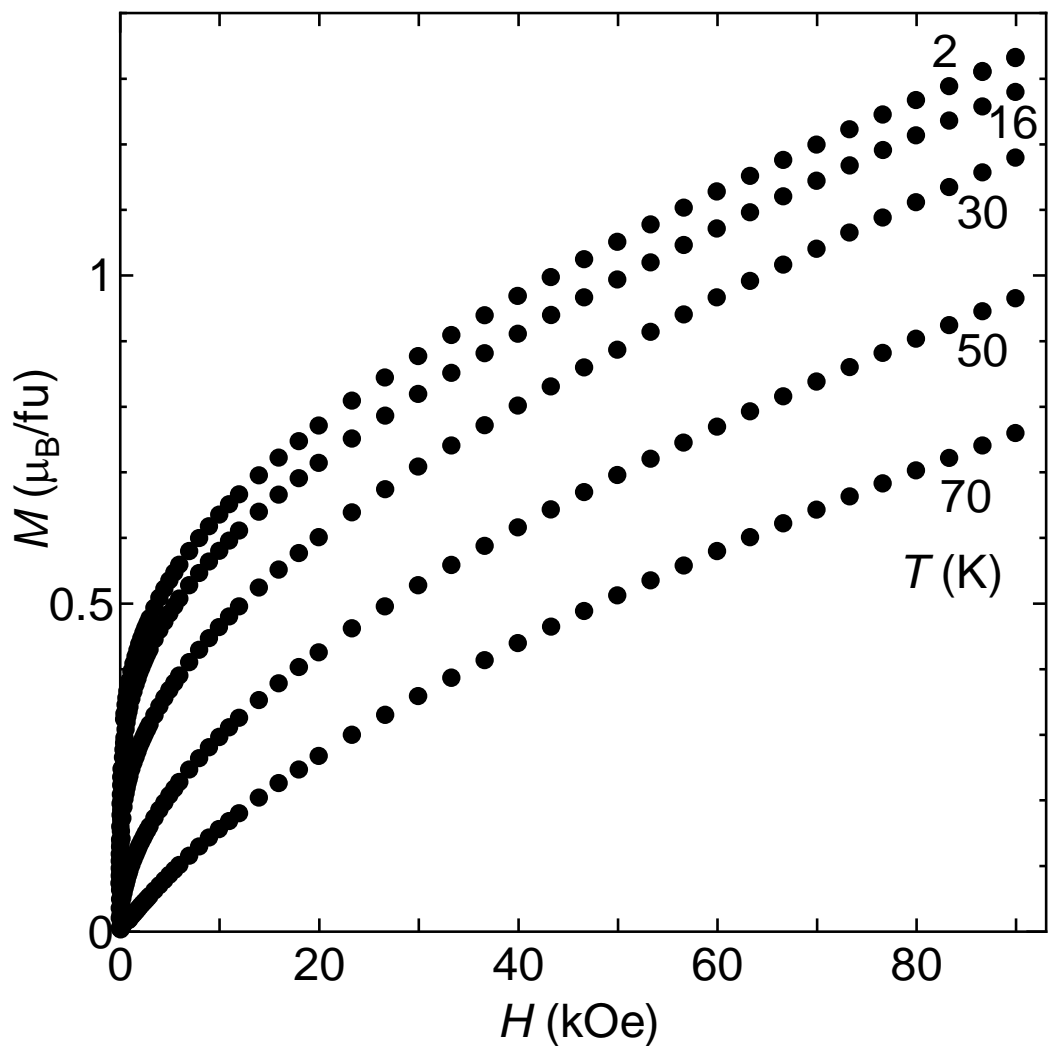

Fig.2, T. Furubayashi

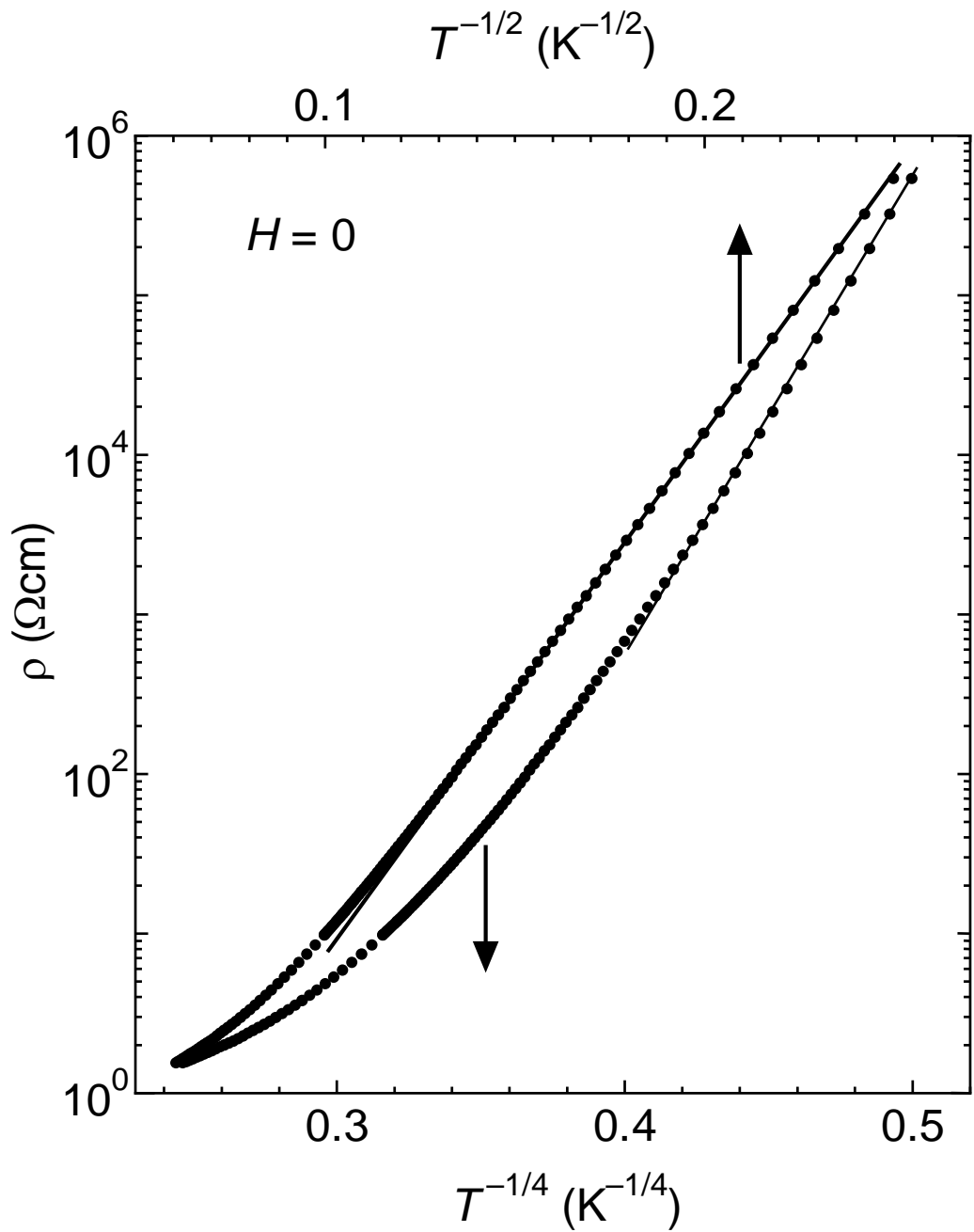

Fig.3, T. Furubayashi

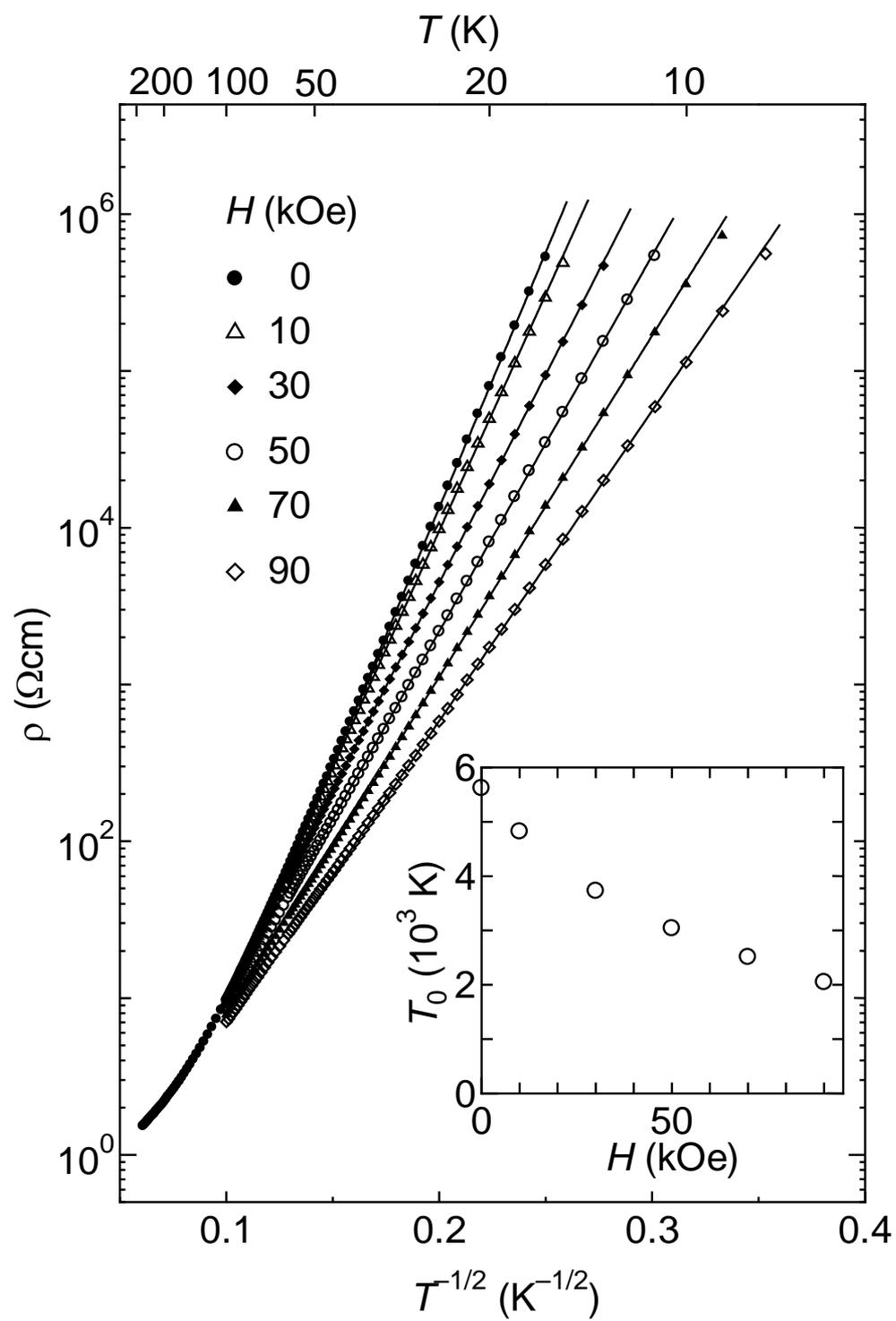

Fig.4, T. Furubayashi

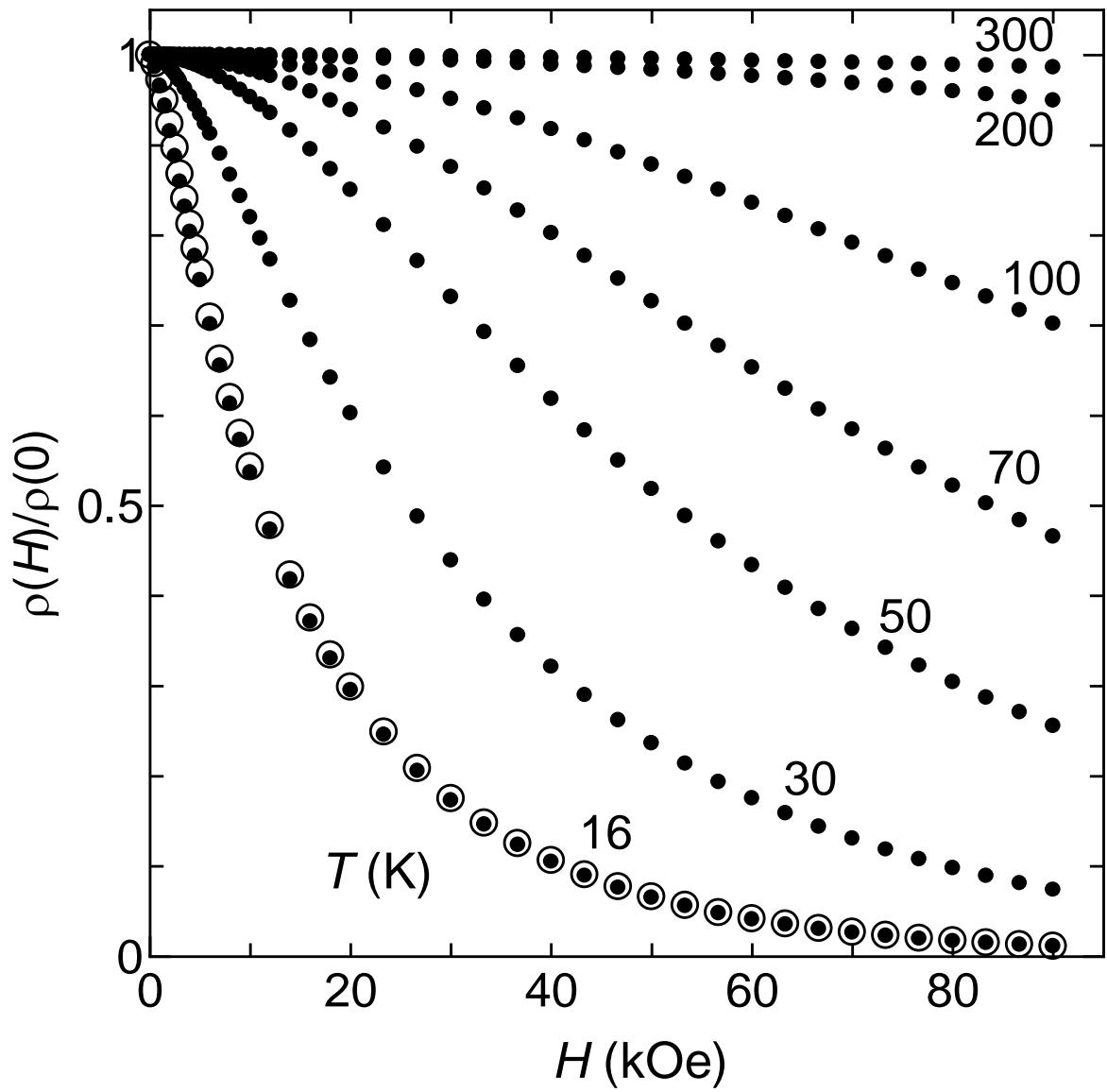

Fig.5, T. Furubayashi